\documentclass[prb,twocolumn,amsmath,amssymb,showpacs]{revtex4}
\usepackage[usenames]{color}
\usepackage[dvipdfm]{graphicx}
\usepackage{subfigure,wrapfig}
\usepackage{dcolumn}% Align table columns on decimal point
\usepackage{bm}% bold math
\usepackage{txfonts}

\begin{document}
\title{Direct observation of correlation time of dynamic nuclear polarization in single quantum dots}
\author{R.\ Kaji}
\email{r-kaji@eng.hokudai.ac.jp}
	\affiliation{Department of Applied Physics, Hokkaido University, N13 W8, Kitaku, Sapporo 060-8628, Japan}

\author{H.\ Sasakura}
	\affiliation{Research Institute for Electronic Science, Hokkaido University,  N21 W10, Kitaku, Sapporo 001-0021, Japan}
	
\author{S.\ Adachi}
	\affiliation{Department of Applied Physics, Hokkaido University, N13 W8, Kitaku, Sapporo 060-8628, Japan}

\author{S.\ Muto}
	\affiliation{Department of Applied Physics, Hokkaido University, N13 W8, Kitaku, Sapporo 060-8628, Japan}
%%%\altaffiliation[Also at ]{ }%Lines break automatically or can be forced with \\

\date{\today}% It is always \today, today,
             %  but any date may be explicitly specified

%%%%%%%%%%%%%%%%%%%%%%%%%%%%%%%%%%%%%%
\begin{abstract}
The spin interaction between an electron and nuclei was investigated optically in a single self-assembled InAlAs quantum dot (QD). In spin dynamics, the correlation time of the coupled electron-nuclear spin system and the electron spin relaxation time play a crucial role. We examined on a positively charged exciton in a QD to evaluate these key time constants directly via the temporal evolution
measurements of the Overhauser shift and the degree of circular polarization. In addition, the validity of our used spin dynamics model was discussed in the context of the experimentally obtained key parameters. 
\end{abstract}
%%%%%%%%%%%%%%%%%%%%%%%%%%%%%%%%%%%%%%
% PACS, the Physics and Astronomy
                             % Classification Scheme.
%\keywords{Zeeman splitting, exciton, quantum dots, hyperfine interaction}
%Use showkeys class option if keyword
                              %display desired
\maketitle

\section{Introduction}
The hyperfine interaction in semiconductor quantum dots (QDs) is enhanced owing to the strong 3D confinement of the electron wave function; consequently, this has attracted considerable attention from the fundamental and practical points of view. The sophisticated control of nuclear spin polarization (NSP) is required for fascinating applications such as a long-lived memory at an atomic level~\cite{Taylor03} and qubit conversion between an electron spin and a photon~\cite{Muto05}. In semiconductor QDs, the enhanced hyperfine interaction provides the possibility of polarizing nuclear spins (n-spins) in one direction with the optically selective excitation of the electron spin (e-spin). In fact, a large NSP of up to 30$-$60$\%$ was observed recently in interface GaAs QDs~\cite{Gammon01}, self-assembled InAlAs QDs~\cite{Yokoi05, Kaji07, Kaji08}, In(Ga)As QDs~\cite{Braun06, Tartakovskii07, Maletinsky07}, and InP QDs~\cite{Chekhovich10}. In these QDs, the confined electron is subject to a large nuclear field (Overhauser field: $B_{\rm N}$) up to several Tesla. In addition, a fundamental interest is the knowledge of the decay of the e-spin polarization induced by the random fluctuation of the Overhauser field. This fluctuation induces an additional e-spin precession around the effective magnetic field, and it imposes an inevitable contribution to the e-spin relaxation and decoherence through the transverse and longitudinal components of $B_{\rm N}$ fluctuation, as predicted in a previous work~\cite{Merkulov02}.  

From these points of view, it is necessary to examine the spin dynamics of a coupled electron-nuclei (e-n) system that is well isolated in a QD. In the framework of the current semiclassical dynamics model of NSP formation~\cite{Kaji08, Kaji_Thesis}, the following two physical quantities play a significant role: the correlation time of the coupled e-n spin system and the e-spin relaxation time. 

The correlation time $\tau_{\rm c}$ indicates the characteristic time during which the randomly fluctuating effective field is considered to be constant in magnitude and direction according to the traditional spin relaxation theory~\cite{OptOrientation, SpinPhysics}, and it induces a homogeneous broadening $\left( 2 \hbar / \tau_{\rm c} \right)$ of the target e-n levels. Since the NSP formation rate is very sensitive to the degree of energy mismatch in the e-n spin flip-flop process, which is determined by the splitting and the broadening of the corresponding e-n energy levels, $\tau_{\rm c}$ has a crucial influence on the NSP dynamics. Despite its importance, this key quantity has been generally used as a fitting parameter in the model calculations to reproduce the observed results, and the direct experimental estimation of $\tau_{\rm c}$ has not been reported thus far in any QD materials. Meanwhile, the e-spin relaxation time $\tau_{\rm s}$ also affects NSP formation by changing the longitudinal component of the e-spin polarization. As mentioned in a later section, the e-spin relaxation rate ($1/\tau_{\rm s}$) can generally be expressed as a Lorentzian function of the effective magnetic field, and it is influenced by the aforementioned correlation time through its width of $2\hbar /\tau_{\rm c}$. Accordingly, we measure the e-spin relaxation time at a zero effective field $\tau_{\rm{s}0}$, which gives the maximum amplitude of $1/\tau_{\rm s}$; this amplitude can be evaluated independently of $\tau_{\rm c}$.

A secondary interest is the possibility of the measurement of NSP in a QD structure through the degree of circular polarization (DCP) of positively charged exciton emissions. The DCP of the time-integrated photoluminescence (PL) has been used as a powerful tool to detect NSP or the Overhauser shift (OHS), which is the energy shift in the electronic level induced by $B_{\rm N}$, in bulk and quantum well structures for a long time. However, the method to probe NSP in single QDs has been limited only to the change in the energy splitting of the PL lines; this is the simplest way to evaluated OHS, but its accuracy has been limited by the spectral resolution of the experimental setup. In the coupled e-n system, by using the DCP of positively charged exciton emission, which directly reflects the e-spin polarization, it may be possible to follow not only the e-spin but n-spin dynamics, and the study of the coupled e-n system may act as a tool for sensitive measurements of QD-NSP.
%A secondary interest is the possibility of the measurement of NSP in a QD structure through the degree of circular polarization (DCP) of positively charged exciton emissions. Although the DCP of the time-integrated photoluminescence (PL) has been used as a powerful tool to detect NSP in bulk and quantum well structures for a long time, the method to probe NSP in QDs has been limited only to the Overhauser shift (OHS), which is the energy shift in the electronic level induced by $B_{\rm N}$. OHS makes it possible to evaluate NSP directly; however, its accuracy has been limited by the spectral resolution of the experimental setup. In the coupled e-n system, by using the DCP of positively charged exciton emission, which directly reflects the e-spin polarization, it may be possible to follow not only the e-spin but n-spin dynamics, and the study of the coupled e-n system may act as a tool for sensitive measurements of QD-NSP.

%\medskip
In this study, we investigated the e-n spin dynamics in QD structures by using the DCP of the positively charged exciton (X$^{+}$). The DCP of X$^{+}$ PL, which is mainly determined by the e-spin polarization, changed in synchronization with the OHS or the energy splitting of the e-spin levels, and this phenomenon provides the possibility of the sensitive probing of the QD-NSP. By taking advantage of this feature, the key quantities ($\tau_{\rm c}$ and $\tau_{\rm{s0}}$) were evaluated directly from the experimental data. In addition, we extended the dynamic model of NSP by including the dynamics of the X$^{+}$ state, and we confirmed the validity of our model by comparing the time-resolved OHS and DCP measurements with the calculated results.

\section{sample and setup}
Self-assembled In$_{0.75}$Al$_{0.25}$As QDs, which were embedded with an Al$_{0.3}$Ga$_{0.7}$As layer grown on an undoped (100) GaAs substrate by molecular beam epitaxy, were used in the experiments. Assuming a lens-shaped QD with a typical diameter of $\sim$20 nm and a height of $\sim$4 nm derived from atomic force microscopy (AFM) measurements, the number of nuclei in a single QD was estimated roughly to be $\sim$3$\times 10^4$. Micro-PL measurements were performed at 6 K under longitudinal magnetic fields ($B_{\rm z}$) of up to 5 T. A cw-Ti:sapphire laser of $\sim$728 nm, which provides the transition energy at the wetting layer of InAlAs QDs, was employed to illuminate the QD sample. The QD emission spectra were detected using a triple grating spectrometer and a liquid N$_2$-cooled Si-CCD detector. Though the energy resolution of our setup was $\sim$12 $\mu$eV, it can be improved to 5 $\mu$eV by spectral fitting. The polarization of the excitation light was controlled using an electro-optic modulator (EOM), and the Zeeman splitting energy and the DCP of a target single QD spectra were evaluated.

\section{Results and Discussion}

\subsection{Electron and nuclear spin polarization in a coupled system}

\begin{figure}[t]
  \begin{center}
    \includegraphics[width=240pt]{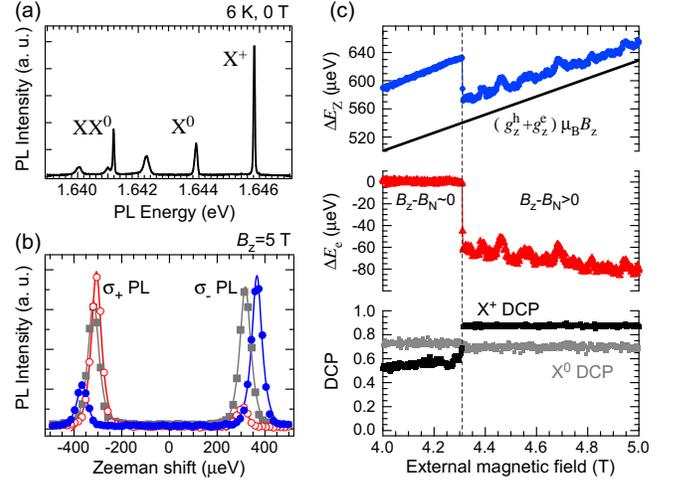}
\caption{(a) PL spectra from the target single QD at zero magnetic field. (b) PL spectra of X$^{+}$ state at $B_{\rm z}=$5 T with linearly (gray squares) and circularly ($\sigma_{+}$/$\sigma_{-}$: open/solid circles) polarized excitation. The $\sigma_{-(+)}$ PL component is positioned at the higher (lower) energy side. The difference between the Zeeman splitting $\Delta E_{\rm Z}$ for the circularly and linearly polarized excitations is defined as the Overhauser shift ($\Delta E_{\rm{OHS}}$), and it is evaluated as $\Delta E_{\rm{OHS}}=$98 $\mu$eV ($-$21 $\mu$eV) for $\sigma_{-}$ ($\sigma_{+}$) excitation. (c) $B_{\rm z}$ dependences of $\Delta E_{\rm Z}$ of the X$^{+}$ PL line  (upper panel), the energy splitting of the electronic level $\Delta E_{\rm e}$ (middle panel), and the DCPs of X$^{+}$ and X$^{0}$ (denoted by the black and gray symbols in lower panel). The solid line in the upper panel is the calculated $\Delta E_{\rm Z}$ at $B_{\rm N}=$0.} 
\label{Fig1}
    \end{center}
\end{figure}

First, we investigate the availability of X$^{+}$ DCP as a powerful measure of the electron and nuclear spin polarization in a QD. Fig.~\ref{Fig1}(a) shows the PL spectra obtained from the target single QD under a zero magnetic field. Regardless of the undoped sample we used, the PL spectra from the various charge states that originated from the same single QD were observed. The charge states of three peaks with high intensities were assigned to a neutral biexciton (XX$^{0}$), a neutral exciton (X$^{0}$), and a positively charged exciton (X$^{+}$) from the lower energy side, respectively. Since X$^{+}$ had the strongest PL intensity in our QD sample and because there were no dark exciton states present, it is expected to be the dominant contributor in NSP formation.

Fig.~\ref{Fig1}(b) depicts the X$^{+}$ PL spectra at $B_{\rm z}=$5 T for the linearly (denoted by grey squares) and the circularly  ($\sigma_{+}$ and $\sigma_{-}$: denoted by open and solid circles) polarized excitations. In the X$^{+}$ states composed of the spin-paired two holes and an electron, the exchange interactions between the electron and hole spins play no role, and the energy splitting of the PL lines ($\Delta E_{\rm Z}$) is determined solely by the Zeeman interaction of the spins with the (effective) magnetic fields. Under this condition, $\Delta E_{\rm Z}$ can be expressed as follows: $\Delta E_{\rm Z} = g_{\rm z}^{\rm h} \mu_{\rm B} B_{\rm z} + g_{\rm z}^{\rm e} \mu_{\rm B} \left( B_{\rm z} \pm B_{\rm N} \right)$, where $g_{\rm z}^{\rm{e} \left( \rm h \right)}$ denotes the electron (hole) g-factor in the growth direction, $\mu_{\rm B}$ denotes the Bohr magneton, and $B_{\rm N}$ denotes the Overhauser field. 
Since the hole spin has a low probability of existence at the nucleus site, the effect of $B_{\rm N}$ on the hole spin could be neglected except for the special case~\cite{Ebel09}. Under a large $B_{\rm z}$ of a few Tesla, $B_{\rm N}$ manifests itself as an OHS defined as $\Delta E_{\rm{OHS}} =g_{\rm z}^{\rm e} \mu_{\rm B} B_{\rm N}$. Since $B_{\rm N}$ is essentially zero for the linearly polarized excitation, OHS is deduced from the difference between $\Delta E_{\rm Z}$ for the circularly and linearly polarized excitations, and it is evaluated in Fig.~\ref{Fig1}(b) as $\Delta E_{\rm{OHS}}=$98 $\mu$eV ($-$21 $\mu$eV) with $\sigma_{-}$ ($\sigma_{+}$) excitation. As per our definition, the $\sigma_{-}$ ($\sigma_{+}$) excitation generates $B_{\rm N}$ in the opposite (same) direction to $B_{\rm z}$, and it induces an apparent increase (decrease) in $\Delta E_{\rm Z}$ because of the relation $g_{\rm z}^{\rm e} \cdot g_{\rm z}^{\rm h} <$0.
Hereafter, we focus on the $\sigma_{-}$ case where the compensation of $B_{\rm z}$ via $B_{\rm N}$ is achieved; consequently, the bistabilities of NSP have been observed for external parameters such as the excitation power~\cite{Tartakovskii07}, excitation polarization~\cite{Braun06}, and external magnetic field~\cite{Maletinsky07}. 

Figure~\ref{Fig1}(c) summarizes the effects of NSP on the X$^{+}$ PL observed in the $B_{\rm z}$ dependence measurement. In the experiment, the excitation polarization was fixed at $\sigma_{-}$, and the external field was swept from 4.0 T to 5.0 T with a sweeping rate of 0.11 T/min. The symbols and the solid line in the upper panel indicate the observed $\Delta E_{\rm Z}$ and the calculated $\Delta E_{\rm Z}$ under the condition when $B_{\rm N}=$0. The difference from the zero $B_{\rm N}$ line is the OHS at each $B_{\rm z}$. As can be clearly observed, an abrupt decrease in $\Delta E_{\rm Z}$ was observed at $B_{\rm z}=$4.31 T owing to the bistable nature of NSP. In order to measure the degree of $B_{\rm z}$ compensation via $B_{\rm N}$, we introduce the effective magnetic field as experienced by e-spin; $B_{\rm{eff}}$($= B_{\rm z}- B_{\rm N}$). By using the previously obtained values of $g_{\rm z}^{\rm h}=$$+$2.54 and $g_{\rm z}^{\rm e}=$$-$0.37~\cite{Kaji07}, we deduced the electronic splitting energy $\Delta E_{\rm e}= g_{\rm z}^{\rm e} \mu_{\rm B} B_{\rm{eff}}$, as shown in the middle panel of Fig.~\ref{Fig1}(c). In the region where $B_{\rm z} <$4.31 T, the absolute value of $\Delta E_{\rm e}$ nearly reduces to zero, and the Overhauser field fully compensates for the external field. With increasing $B_{\rm z}$, the magnitude of $B_{\rm N}$ shows a clear reduction and $| \Delta E_{\rm e}|$ increases abruptly. 

Here, we focus on the DCP of the X$^{+}$ PL (the lower panel of Fig.~\ref{Fig1}(c)). In this work, the DCP is defined as $\left( I^{-}- I^{+} \right)/ \left( I^{-}+ I^{+} \right)$ ($I^{+ (-)}$ denotes the integrated PL intensity of the $\sigma_{+(-)}$ component). It is noteworthy that the DCP of X$^{+}$ PL shows a clear jump from $\sim$0.6 to $\sim$0.9; this transition synchronizes with the decrease in $\Delta E_{\rm e}$. As mentioned above, the DCP of X$^{+}$ is essentially determined solely by the e-spin polarization $\left< S_{\rm z} \right>$, and it can be expressed as DCP$=$2$\left< S_{\rm z} \right>$. Accordingly, a high (low) value of DCP indicates a small (large) reduction in e-spin polarization (i.e., e-spin relaxation). Since the e-spin relaxation rate depends on $|\Delta E_{\rm e}|$ (see Eq.~\ref{Eq1} in the next section), the change in the X$^{+}$ DCP presents the possibility of a direct measurements of the electron and nuclear spin polarizations in a coupled e-n system.

It is noteworthy that the DCP observed in the other charge states show different behaviors. The OHS observed in the X$^{0}$ and XX$^{0}$ PLs show changes similar to that observed in X$^{+}$; this is one of the evidences that these PL lines originate from the same QD~\cite{Sasakura08}. In contrast, the tendencies of DCP are quite different for these other exciton complex peaks. The X$^{0}$ DCP stays constant ($\sim$0.7), thereby signifying independence from $\Delta E_{\rm e}$, as shown in the lower panel of Fig.~\ref{Fig1}(c) (denoted by the gray symbols). This can be attributed to the contribution of the unpolarized X$^{0}$ supplied from the XX$^{0}$. XX$^{0}$ decays to X$^{0}$ by emitting $\sigma_{+}$ and $\sigma_{-}$ photons with identical probabilities, and therefore, XX$^{0}$ DCP is basically zero (not shown here). The DCP of X$^{0}$ is approximately calculated as  $[(n+n/4)-n/4]/[(n+n/4)+n/4] \sim 0.67$, if QDs are excited under the power at which $n$ electron-hole pairs (X$^{0}$) are generated in each QD (in this case, XX$^{0}$/X$^{0}$=$1/2$ according to Poisson statics). Further studies in this direction requires a close examination of the fine structures of the exciton levels because of the electron-hole exchange interaction.

\subsection{Experimental estimation of correlation time and electron spin relaxation time}

\begin{figure}[t]
  \begin{center}
    \includegraphics[width=200pt]{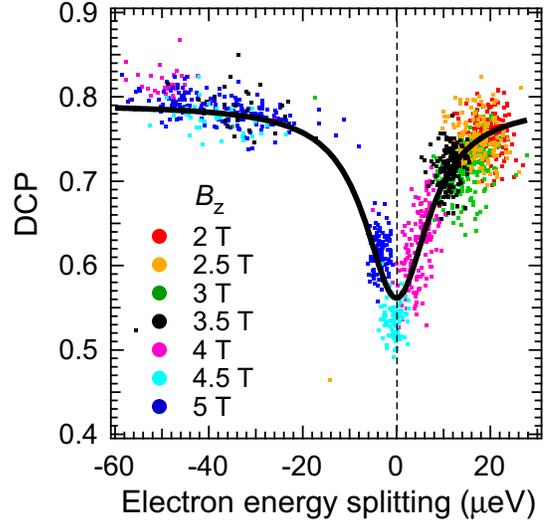}
\caption{X$^{+}$ DCPs as a function of the electronic energy splitting at the different external field. Symbols and colors indicate the experimental data and corresponding $B_{\rm z}$. Each of the data values was obtained from the time-resolved measurements of OHS and DCP. The absence of the data points around $\Delta E_{\rm e}$$\sim$$-15 \ \mu$eV is attributed to the abrupt changes in OHS and DCP. The solid curve represents the Lorentzian fitting with a width of $\sim$15 $\mu$eV and an offset of $\sim$0.8.} 
\label{Fig2}
    \end{center}
\end{figure}

We estimated the key quantities ($\tau_{\rm c}$ and $\tau_{\rm{s0}}$) in the e-n spin dynamics from the experimental data. The e-spin relaxation rate under the effective magnetic field in frequency unit $\Omega_{\rm{e}}$($=\Delta E_{\rm e}/ \hbar$) can generally expressed as~\cite{OptOrientation, SpinPhysics}  
\begin{equation}
\frac{1}{\tau_{\rm s}} = \frac{1}{\tau_{\rm{s0}}} \cdot \frac{1}{1+ \left( \Omega_{\rm e} \tau_{\rm c} \right)^{2}}.
 \label{Eq1}
\end{equation}
This equation represents a Lorentzian shape with a full width at the half maximum (FWHM) of $2 \hbar /\tau_{\rm c}$ in energy and an amplitude of $1/\tau_{\rm{s0}}$. Here, $\tau_{\rm c}$ and $\tau_{\rm{s0}}$ correspond to the e-n correlation time and the e-spin relaxation time at $\Omega_{\rm e} =$0, respectively. As mentioned above, the e-spin relaxation rate appears directly in the reduction in X$^{+}$ DCP, and these key quantities can be obtained from the fitting with the inverse of Eq.~\ref{Eq1}.

Figure~\ref{Fig2} shows the DCPs as a function of $\Delta E_{\rm e}$ at different values of external field strength (2 T $\le B_{\rm z} \le$ 5 T). The data were obtained from the time-resolved measurements, as shown in a later section of the paper (Fig.~\ref{Fig3}(b)). As clearly shown, a definite dip is observed at $\Delta E_{\rm e} \simeq 0$. The Lorentzian fitting (FWHM$\sim$15 $\mu$eV) with an offset ($\sim$0.8) depicted by a solid curve in the figure was able to reproduce the entire experimental data. It should be noted that the X$^{+}$ DCPs at the different $B_{\rm z}$ values depict a unique curve, and this fact justifies the assumption that $\tau_{\rm c}$ and $\tau_{\rm{s0}}$ are independent of $B_{\rm z}$.

First, we estimate the e-n correlation time $\tau_{\rm c}$. By using the relation of FWHM$=2 \hbar / \tau_{\rm c}$, we evaluated the correlation time to be $\tau_{\rm c}\sim$80 ps. The obtained $\tau_{\rm c}$ coincides with the values that Maletinsky \textit{et al}.~\cite{Maletinsky07} (35 ps) and Braun \textit{et al}.~\cite{Braun06} (50 ps) used in the calculations to reproduce their observations in single In(Ga)As QDs. To the best of our knowledge, an experimental estimation of this correlation time has not been reported thus far although $\tau_{\rm c}$ of several tens of picoseconds has been used in the data fitting. 
In addition, the value is in the same magnitude as the X$^{+}$ decoherence time ($\sim $43 ps) that was measured by Fourier spectroscopy of the same InAlAs QD~\cite{Adachi07}.

Secondly, we focus on the e-spin relaxation time under a zero effective magnetic field. This characteristic time can be expressed as $\tau_{\rm{s0}} = {\tau_{\rm R}}/\left[{S_{\rm{op}}/S_{\rm z} \left( 0 \right) -1}\right]$~\cite{OptOrientation}, where $S_{\rm{op}}$ denotes the initial e-spin polarization injected into the QD ground state, and $S_{\rm z} \left( 0 \right)$ is one when $B_{\rm{eff}}$=0. Here, $\tau_{\rm R}$ denotes the recombination time, and it is found to be $\sim$0.75 ns by other independent time-resolved measurements. From the DCP at $\Delta E_{\rm e}$=0, $S_{\rm z} \left( 0 \right)$ is evaluated to be $\sim$0.6. 
In contrast, precise evaluation of $S_{\rm{op}}$ is difficult because the e-spin polarization created with the $\sigma_{-}$ excitation is lost partially during the energy relaxation process from the wetting layer to the QD ground state. 
For simplicity, $S_{\rm{op}}$ is replaced by the offset value of the DCP curve ($\sim$0.8) because the e-spin relaxation process is strongly suppressed in the region of large $B_{\rm{eff}}$ according to Eq.~\ref{Eq1}. By using these values, we evaluated the e-spin relaxation time under $\Delta E_{\rm e}$$=$0 to be $\tau_{\rm{s0}}$$\sim$3$\tau_{\rm R}$. This value is in good agreement with the exciton spin relaxation time obtained in other measurements for resonant and nonresonant excitations~\cite{Watanuki05, Kumano06, Kaji09}. 

Thus far, we were able to estimate $\tau_{\rm c}$ and $\tau_{\rm{s0}}$ directly from the experimental data. Next, it is required to determine the factors influencing these characteristic times. Here, we consider the effects of the randomly fluctuating Overhauser field induced by the n-spin ensemble~\cite{Maletinsky_thesis}. The random fluctuation of $B_{\rm N}$ (denoted by $\Delta B_{\rm N}$) induces additional e-spin precession, and the memory of the e-spin polarization is lost from the initial value. According to the traditional spin precession model, the longitudinal component of $\Delta B_{\rm N}$ ($\Delta B_{\rm{N}, \|}$) induces a loss in the transverse component of e-spin polarization (\textit{i.e.}, decoherence). On the other hand, the transverse component of $\Delta B_{\rm N}$ ($\Delta B_{\rm{N}, \bot}$) gives rise to a loss in the longitudinal e-spin polarization (\textit{i.e.}, relaxation). Assuming a random variable in NSP with a Gaussian distribution of a width $\sqrt{N}$ ($N$: the number of the nuclei in a QD), the fluctuation of $B_{\rm N}$ can be estimated as $\Delta B_{\rm N} \cong A/ \sqrt{N} g^{\rm e} \mu_{\rm B}$ ($g^{\rm e}$: an isotropic electron g-factor, $A$: the hyperfine coupling constant). By using the typical values for a InAlAs QD ($g^{\rm e}$, $A$, $N$) = ($-$0.37, 50 $\mu$eV, 3$\times$10$^{4}$), we can roughly estimate the fluctuation of $B_{\rm N}$ to be $\Delta B_{\rm N} \approx$15 mT. The corresponding decoherence time ($T_{2, \rm{n}}^{*}$) and the relaxation time $\tau_{\rm{en0}}$ are given as the functions of $\Delta B_{\rm N}$. 

To begin with, we compare $T_{2, \rm{n}}^{*}$ and the observed correlation time $\tau_{\rm c}$~\cite{Edit2}. 
%Although $\tau_{\rm c}$ is limited by the shortest decoherence time of the coupled e-n spin system, the decoherence time of the decoupled e-spin system is thought to determine $\tau_{\rm c}$ mainly attributed to the long decoherence time of the n-spin system~\cite{Edit2}. 
The e-spin decoherence time induced by $\Delta B_{\rm N, \|}$ can be expressed as $T_{2, \rm{n}}^{*} = \hbar / \left[ g_{\rm z}^{\rm e} \mu_{\rm B} \sqrt{2 \Delta B_{\rm N}^{2} /3} \right]$,~\cite{Merkulov02_A} and it is estimated to have an approximate value of $\sim$3 ns. The fact that the estimated $T_{2, \rm{n}}^{*}$ is fairly longer than the observed $\tau_{\rm c}$ may indicate the presence of another scattering process which is responsible for the shorter decoherence time $T_{2}^{*}$. In our assumption, $1/ \tau_{\rm c}$ is given as a linear coupling of $1/T_{2, \rm{n}}^{*}$ and $1/ T_{2}^{*}$ and is dominantly determined by $1/T_{2}^{*}$. 
One of the plausible causes for $1/T_{2}^{*}$ is the charge fluctuation in the QD region. Lai \textit{et al}. have mentioned the effect of the e-spin tunneling between the QD and the n-doped layer in their charge-tunable QD structure~\cite{Lai06}. 
Although our QD sample has no diode structure and there is no interaction with the electrode, unlike their charge-tunable QD sample, similar phenomena such as carrier tunneling between QD and its surroundings occur even in our sample; consequently various types of the charge states appear during the exposure time of the CCD detector ($\sim$0.1-1 s), as shown in Fig.~\ref{Fig1}(a). The changes of the charge state
 may interrupt the e-spin precession, and it may affect the correlation time of an e-n spin system. 
%Although our QD sample has no diode structure and there is no interaction with the electrode through phenomena such as e-spin tunneling, various types of the charge states appear during the exposure time of the CCD detector ($\sim$0.1-1 s), as shown in Fig.~\ref{Fig1}(a). 
%This charge fluctuation may interrupt the e-spin precession, and it may affect the correlation time of an e-n spin system. 
%Furthermore, excess carriers existing in the QD surrounding may contribute to $1/T_{2}^{*}$ via Coulomb interaction. 
Further 0research is required regarding the origin of $1/T_{2}^{*}$. 

Next, we compare $\tau_{\rm{en0}}$ and the observed e-spin relaxation time under $B_{\rm{eff}}=$0. 
Assuming an uniform electron wave function in the QD region, the e-spin relaxation time induced by $\Delta B_{\rm{N}, \bot}$ under $B_{\rm eff}$=0, which is given as $\tau_{\rm{en0}}= \left[ N \tau_{\rm c} \left( A/ N \hbar \right)^{2} \right]^{-1}$, to be $\sim$50 ns~\cite{Edit3}.
%By assuming an uniform electron wave function in the QD region, the e-spin relaxation time induced by $\Delta B_{\rm{N}, \bot}$ under $B_{\rm eff}$=0 is given as $\tau_{\rm{en0}}= \left[ N \tau_{\rm c} \left( A/ N \hbar \right)^{2} \right]^{-1}$ by the motional narrowing theory~\cite{OptOrientation,SpinPhysics}, and it is estimated to be approximately $\sim$50 ns~\cite{Edit3}. 
Since the estimated $\tau_{\rm{en0}}$ is one order of magnitude longer than the observed $\tau_{\rm{s0}}$, we introduce the e-spin flip term ($1/\tau_{\rm{e0}}$) separately from the e-n flip-flop term ($1/\tau_{\rm{en0}}$), and the total e-spin relaxation rate obtained from the measurements is assumed to be $1/\tau_{\rm{s0}} = 1/\tau_{\rm{en0}} +1/\tau_{\rm{e0}}$. In our assumption, these two spin relaxation processes have the same $\Omega_{\rm e}$ dependency as that represented by a common Lorentzian function. Note that the large reduction in the DCP shown in Fig.~\ref{Fig2} cannot be reproduced without the introduction of $1/\tau_{\rm{e0}}$.

In this section, the e-n correlation time and the e-spin relaxation time at $B_{\rm{eff}}=$0, two key quantities in e-n spin dynamics, were evaluated from the experimental data. Since the estimated spin decoherence and relaxation time induced by the random fluctuation of $B_{\rm N}$ are of fairly longer duration than the values obtained from the measurements, other scattering processes are required to explain these durations. Although the source that decides $\tau_{\rm c}$ and $\tau_{\rm{s0}}$ has not yet been identified at the present stage, a direct evaluation provides the valuable information for the modeling of e-n spin dynamics, as discussed in the next section.

\subsection{Dynamics model of coupled electron-nuclear spin system}

\begin{figure}[t]
  \begin{center}
    \includegraphics[width=240pt]{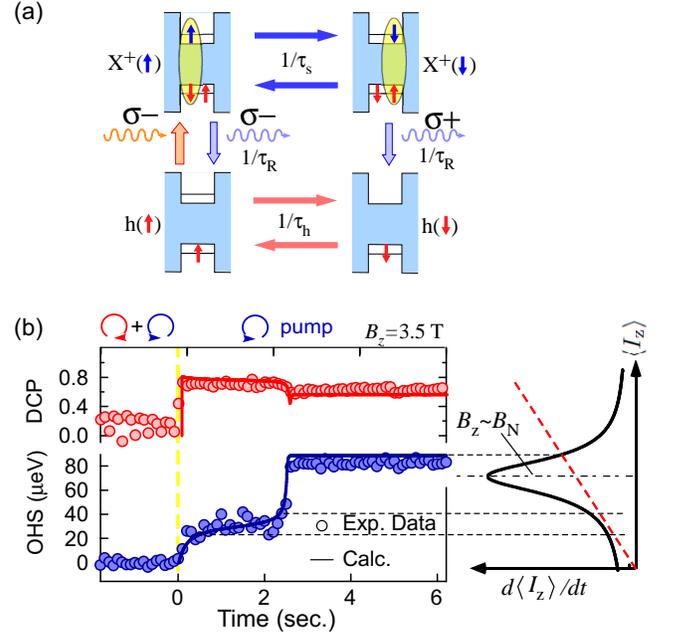}
\caption{(a) Current dynamics model of e-spin system including X$^{+}$ and the single hole states, and the corresponding PL polarizations. (b) Transient evolutions of DCP and OHS at $B_{\rm z}=$3.5 T. The excitation polarizations are depicted in the upper side of the panel. The solid curves are the calculated results in the coupled e-n system, and they were able to reproduce all the behaviors of the experimental results. Right inset indicates a schematic of the n-spin polarization (black solid curve) and depolarization (red dashed line) terms for an explanation of the transient OHS.} 
\label{Fig3}
    \end{center}
\end{figure}

Finally, we test the validity of the dynamics model of the coupled e-n spin system. The temporal evolution of the mean NSP $\left< I_{\rm z} \right>$ is described by the following rate equation~\cite{Abragam}:
\begin{equation}
\frac{d \left< I_{\rm z} \right>}{dt} = \frac{1}{T_{\rm{NF}}} \left[ Q \left( \left< S_{\rm z} \right> -S_{0} \right) -\left< I_{\rm z} \right> \right] -\frac{1}{T_{\rm{ND}}} \left< I_{\rm z} \right>
\label{Eq2}
\end{equation}
where $S_{0}$ denotes the thermal e-spin polarization, $1/T_{\rm{NF}}$ and $1/ T_{\rm{ND}}$ denote the n-spin polarization and depolarization rates~\cite{Edit4}, respectively, and $Q = I \left( I+1 \right)/ \left[ S \left( S+1 \right) \right]$ denotes the momentum conversion coefficient from the e-spin to n-spin system. It is noteworthy that the n-spin polarization rate is also responsible for the e-spin relaxation, and it appears in the e-spin dynamics via $1/\tau_{\rm{en0}} = N /T_{\rm{NF0}}$ ($1/T_{\rm{NF}0}$: the n-spin polarization rate at $B_{\rm{eff}}=$0). 
Although Eq.~\ref{Eq2}, with constant values of the averaged e-spin polarization $\left< S_{\rm z} \right>$, has explained the observed OHS qualitatively in previous studies, the actual $\left< S_{\rm z} \right>$ in the dynamics is expected to change along with the evolution of $\left< I_{\rm z} \right>$. In our model calculation, $\left< S_{\rm z} \right>$, which drags the randomly-oriented n-spin ensemble to the highly polarized state, is determined by the dynamics in the following four states, as shown in Fig.~\ref{Fig3}(a): X$^{+}$ with the spin-up/down electron ($n_{\uparrow}$ and $n_{\downarrow}$: the populations of the corresponding states), and the spin-up/down single hole states (similarly denoted by $n_{\Uparrow}$ and $n_{\Downarrow}$), and $\left< S_{\rm z} \right>$ is given as $\left( n_{\uparrow} - n_{\downarrow} \right) / \left[ 2 \left( n_{\uparrow} + n_{\downarrow} \right) \right]$. These four states are connected with the rates of the optical pumping with $\sigma_{-}$ light, the radiative recombination ($1/\tau_{\rm R}$), and the spin flip of electron and hole ($1/\tau_{\rm s}$ and $1/\tau_{\rm h}$).

The direct observation of the temporal evolutions of OHS and DCP can provide a better understanding of the e-n spin dynamics. Typical transients obtained from the target X$^{+}$ PL at $B_{\rm z}=$3.5 T are shown in Fig.~\ref{Fig3}(b). In order to set the initial NSP to zero, the excitation polarization before the time of origin was modulated between $\sigma_{+}$ and $\sigma_{-}$, with a frequency of 10 Hz. The temporal evolution of OHS (=$A \left< I_{\rm z} \right>$) is explained schematically by the difference between the n-spin polarization (a black solid curve) and the depolarization (a red dashed line) rates, as shown in the right inset of Fig.~\ref{Fig3}(b). After switching to $\sigma_{-}$ excitation, OHS increases gradually in the region of small difference and increases explosively around the peak of the n-spin polarization rate. Under this experimental condition, the OHS jumps clearly to the saturated value within 3 s, and $B_{\rm z}$ compensation via $B_{\rm N}$ is achieved within the homogeneous broadening of the n-spin polarization rate. At the exact moment of the abrupt increase in the OHS, the DCP of the X$^{+}$ PL drops suddenly from 0.8 to 0.7. The solid curves are the calculated results obtained from the abovementioned dynamics model. In the calculation, the key parameters ($\tau_{\rm c}$ and $\tau_{\rm{s0}}$) were in the same order as the experimentally evaluated values reported in a previous section of the paper. The fact that the calculation could reproduce the observed DCP as well as OHS shows the validity of our dynamics model.

%\smallskip
\section{Conclusion}
In conclusion, we investigated the spin dynamics of the coupled electron-nuclear spin system in a single InAlAs QD. The DCP of X$^{+}$ PL, which is basically determined by e-spin polarization, showed synchronized changes with the electronic energy splitting, and this fact offers the possibility of NSP probing via X$^{+}$ DCP in a QD structure. 
By taking advantage of this feature, the correlation time of the e-n spin system and the e-spin relaxation time, which play a crucial role in spin dynamics, were evaluated as $\tau_{\rm c} \sim$80 ps and $\tau_{\rm{s0}} \sim$3$\tau_{\rm R}$, respectively. The experimentally obtained $\tau_{\rm c}$ agrees well with the results obtained by calculations or the decoherence time of X$^{+}$ PL; further, $\tau_{\rm{s0}}$ agrees well with the exciton spin relaxation time obtained from other experiments. Although the definite source of these key quantities were not identified at this stage, a direct estimation from measurements is very important in the characterization of the e-n spin dynamics. 
The spin dynamics model used in this study successfully reproduces the observations of DCP as well as the OHS, and we believe that the model can significantly contribute to the understanding of e-n spin dynamics.


\begin{thebibliography}{00}
\bibitem{Edit2} Although $\tau_{\rm c}$ is limited by the shortest decoherence time of the coupled e-n spin system, the decoherence time of the decoupled e-spin system is thought to determine $\tau_{\rm c}$ mainly. The decoherence time of n-spin ensemble was estimated as the order of hundreds of milliseconds by the n-spin diffusion constant measurements.
\bibitem{Edit3} Strictly speaking, $\tau_{\rm{en}0}$ also includes the factor of finite electron occupancy in QD ($f_{\rm e} \le$1), and $f_{\rm{e}} \tau_{\rm{en}0}$ is used in the calculation. Thus, the estimated $\tau_{\rm{en}0}$ of $\sim$50 ns may give the lower limit of the e-spin relaxation time due to the hyperfine interaction.
\bibitem{Edit4} For simplicity, the magnetic field dependence of $1/T_{\rm{ND}}$ is ignored here and it is not necessary to introduce this dependence at the present stage.
\end{thebibliography}
\end{document}